\documentclass[twocolumn,showpacs,prd]{revtex4}
\usepackage{graphicx}
\usepackage{dcolumn}
\usepackage{bm}
\begin{document}

\title{Electron Capture in a Fully Ionized Plasma}
\author{A. Widom and J. Swain}
\affiliation{Physics Department, Northeastern University, Boston MA USA}
\author{Y.N. Srivastava}
\affiliation{Physics Department, University of Perugia, Perugia IT}

\begin{abstract}
Properties of fully ionized water plasmas are discussed including plasma 
charge density oscillations and the screening of the Coulomb law especially 
in the dilute classical Debye regime. A kinetic model with two charged 
particle scattering events determines the transition rate per unit time 
for electron capture by a nucleus with the resulting nuclear transmutations. 
Two corrections to the recent Maiani {\em et.al.} calculations are made: 
(i) The Debye screening length is only employed within its proper domain 
of validity. (ii) The WKB approximation employed by Maiani in the long 
De Broglie wave length limit is evidently invalid. We replace this incorrect    
approximation with mathematically rigorous Calogero inequalities 
in order to discuss the scattering wave functions. Having made these 
corrections, we find a verification for our previous results based on 
condensed matter electro-weak quantum field theory for nuclear transmutations 
in chemical batteries.
\end{abstract}


\maketitle

\section{Introduction \label{intro}}

In recent years we have been working on electro-weak
interaction inverse beta decay by including including 
electro-magnetic interactions with collective plasma modes 
of motion\cite{widom:2006,widom:2007}. We have applied this 
theory to electron capture in a water plasma to explain observed 
nuclear transmutations on the cathode surface of a chemical 
cell\cite{Cirillo:2012}. While the original theory was formulated 
in terms of electro-weak quantum field theory\cite{Donoghue:1992} 
in a many body context, a reasonable alternative relies on physical 
kinetic plasma 
theory\cite{Landau:1980,Lifshitz:2006,Pitaevskii:2006,Abrikosov:1975} 
to describe a water plasma. The theoretical kinetic model gives rise 
to electron capture rates per unit time per unit cathode surface 
area in a water plasma in agreement with the quantum field theoretical  
model and is in agreement with experiments. 

Objections based on the kinetic model in a {\em cold plasma} were 
raised by the Rome group Ciuchi {\em et.al.}\cite{Ciuchi:2012}. They 
find electron capture rates about two orders of magnitude lower than 
our previous work. The objection was answered by pointing out that 
the water plasma in a chemical cell can light up the laboratory and 
thereby represents a {\em hot plasma}. The {\em hot fully ionized plasma} 
gives rise to an increased electron capture 
rate in agreement with experiment and in agreement with our previous 
results. This has been previously and fully 
discussed\cite{widom:2112,widom:2113}.

Most recently it has been predicted by the Rome group, Maiani 
{\em et.al.}\cite{Maiani:2014}, that a {\em cold plasma} has a higher 
rate of electron capture than does a {\em hot plasma}. This has been 
predicted by Maiani on the basis of (i) the Debye screening of
the attractive Coulomb interactions between the electron and the 
proton and (ii) by the quasi-classical WKB approximation to the s-wave 
electron-proton wave function.  The prediction is in
flagrant disagreement with experiments which exhibit a hot water 
plasma nuclear transmutations and do not exhibit such transmutations 
in a cold plasma. The Maiani computation fails theoretically because 
(i) The Debye screening length is applied in regimes wherein it is  
clearly invalid and (ii) the WKB approximation is applied in the long 
De Broglie wave length regime but in reality the WKB approximation 
is valid only in the short De Broglie wave length regime. One of our 
purposes is to correct the errors made by Maiani. When the properly 
rigorous mathematics is applied we recover our previous and correct 
re3sults. 

In Sec.\ref{tst}, rigorous sum rules for the plasma oscillation 
frequency and the plasma  screening length is reviewed. A general 
thermodynamic expression for the screening length is found in 
Sec.\ref{tsr}. The Debye screening length is then derived in 
Sec.\ref{ds} and the regime of the validity of the Debye theory is 
clearly specified. 

In Sec.\ref{ec} the expression for an electron 
capture transition rate per unit time per nucleus is derived in 
terms of the electron nuclear correlation function 
\begin{equation}
\bar{n}=\left<\sum_{j}\delta \big({\bf R}-{\bf r}_j\big)\right>
\label{intro1}
\end{equation}
describing the density of electrons at positions 
\begin{math} \{{\bf r}_j\} \end{math} sitting right on top of a 
nucleus at position \begin{math} {\bf R} \end{math}. The effects 
of the plasma on electron capture transition rates is described by 
\begin{math} \bar{n} \end{math}. For the case of a water plasma on 
the cathode surface of a chemical cell exhibiting nuclear 
transmutations, the length scales are discussed and estimated  
in Sec.\ref{ls}. In Sec.\ref{ds}, a hot Debye screened plasma result 
is derived for \begin{math} \bar{n} \end{math} equivalent to 
our previous calculations\cite{widom:2112,widom:2113} but in dissagreement 
with  Maiani {\em et.al.}\cite{Maiani:2014} for reasons discussed above.

The Rome group in reality calculates\cite{Maiani:2014} for some 
densities in the quantum degenerate zero temperature regime wherein 
Thomas-Fermi quantum screening plasma replaces classical Debye 
screening plasma. In practical terms, this regime requires the 
solution of the radial s-wave potential scattering equation  
\begin{equation}
-\left(\frac{\hbar ^2}{2\mu }\right)\frac{d^2 u(r)}{dr^2} 
+{\cal U}(r)u(r)=Eu(r), 
\label{intro2}
\end{equation}
wherein \begin{math} {\cal U}(r) \end{math} is the 
two charged particle screened potential. In solving the problem 
for pure s-wave scattering one conventionally takes the limit 
\begin{math} E\to 0 \end{math}. Maiani\cite{Maiani:2014} 
unconventionally employs the WKB method valid only for  
\begin{math} E\to \infty \end{math} wherein phase shifts other 
than s-wave gain importance. These two energy limits are different.
Maiani is in error in his calculation of of s-wave potential scattering 
wave functions.

For completeness of presentation, in Sec.\ref{ssw} we review the 
proper method of computing s-wave scattering wave functions employing 
the mathematically rigorous variable phase formalism\cite{Calogero:1967} 
of Calogero. In the limit \begin{math} E\to 0 \end{math}, one may compute 
the scattering wave functions in terms of the variable scattering length 
as discussed in Sec.\ref{vsl}. The implications for electron capture rates 
is discussed in Sec.\ref{dec}. In the concluding Sec.\ref{conc} a summary 
is given of the results of this work.

\section{Screening Theory \label{tst}}

\begin{figure}
\scalebox {0.6}{\includegraphics{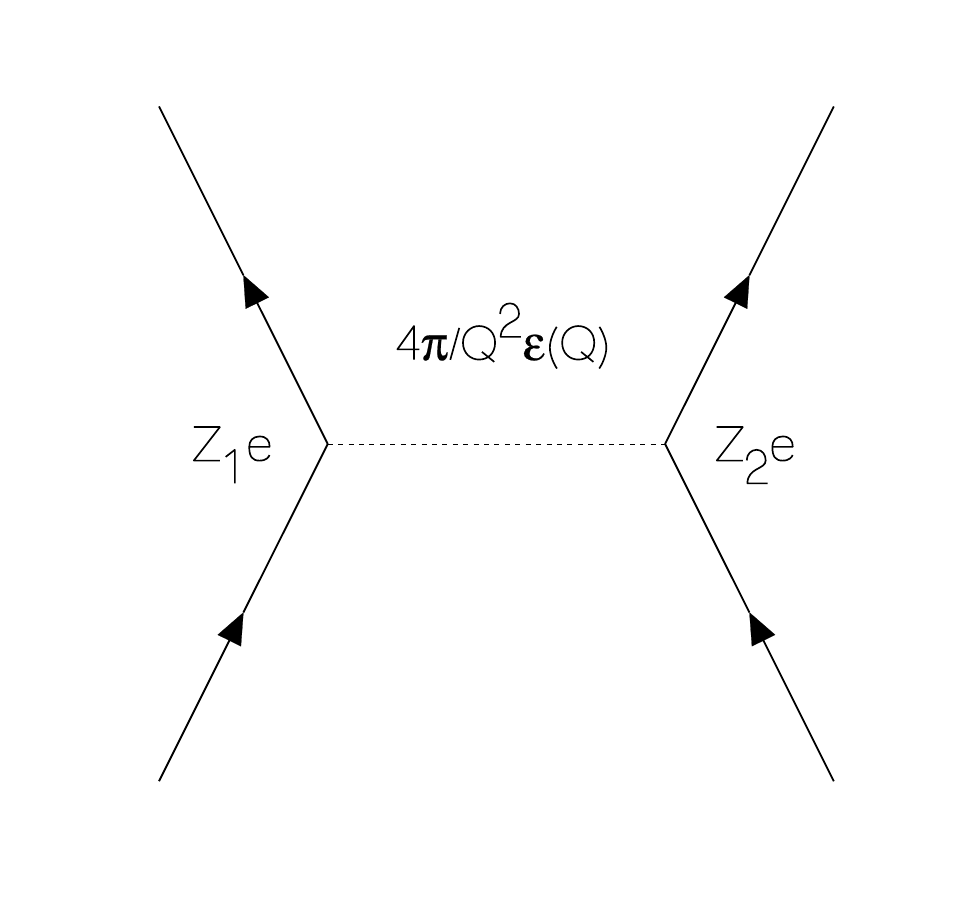}}
\caption{The interaction  
$4\pi e^2 Z_1Z_2/Q^2 \varepsilon (Q)$
determines the screened Coulomb interaction potential 
energy between two charged particles due to one photon 
exchange. The static interaction potential energy is 
given in Eq.(\ref{tst1}).} 
\label{fig1}
\end{figure}

The dielectric response of a plasma to external charge distributions is described by
a wave number \begin{math} {\bf Q}  \end{math} and complex frequency
\begin{math} \zeta  \end{math} dependent dielectric function
\begin{math} \varepsilon (Q,\zeta ) \end{math} in the upper half frequency plane
\begin{math} \Im m\ \zeta > 0  \end{math}. The static dielectric response function
\begin{math} \varepsilon (Q)=\lim_{\omega \to 0} \varepsilon (Q,\omega+i0^+ )\end{math} 
determines the one photon exchange screened Coulomb interaction between two charges 
\begin{math} Z_1e  \end{math} and \begin{math} Z_2e  \end{math},  
\begin{eqnarray}
{\cal U}(r)=Z_1Z_2e^2\int e^{i{\bf Q\cdot r}} 
\left[\frac{4\pi }{Q^2 \varepsilon (Q)} \right]
\frac{d^3{\bf Q}}{(2\pi )^3}\ , 
\nonumber \\ 
{\cal U}(r)= \left(\frac{Z_1Z_2e^2}{r}\right){\cal S}(r),
\label{tst1}
\end{eqnarray}
with a Coulomb law screening function  
\begin{eqnarray}
{\cal S}(r)=\left(\frac{2}{\pi }\right)
\int_0^\infty \sin(Qr)\left[\frac{dQ}{Q\varepsilon (Q)}\right],
\nonumber \\ 
\frac{1}{Q\varepsilon (Q)}=\int_0^\infty \sin(Qr) S(r) dr,
\label{tst2} 
\end{eqnarray}
as shown in FIG. \ref{fig1}.

The screening length \begin{math} \Lambda =\kappa^{-1} \end{math} may be 
defined as the equivalent limits 
\begin{eqnarray}
\varepsilon (Q)=1+\frac{\kappa^2}{Q^2}+\cdots \ \ \ \ (Q\to 0), 
\nonumber \\ 
{\cal S}(r)\to e^{-r/\Lambda}+\cdots  \ \ \ \ (r\to \infty).
\label{tst3} 
\end{eqnarray}
In virtue of Eqs.(\ref{tst2}) and (\ref{tst3}) one may write the formal 
limits    
\begin{eqnarray} 
\kappa^2=\lim_{Q\to 0} Q^2\varepsilon (Q), 
\nonumber \\ 
\frac{1}{\kappa^2}=\Lambda^2=\int_0^\infty rS(r) dr.
\label{tst4}
\end{eqnarray}

If we choose the limit \begin{math} Q\to 0  \end{math} first, then the 
plasma conductivity \begin{math} \sigma (\zeta )  \end{math} is defined 
and obeys a dispersion relation,  
\begin{eqnarray}
\lim_{Q\to 0} \varepsilon(Q,\zeta )=
1+\frac{4\pi i\sigma (\zeta )}{\zeta } 
\nonumber \\ 
\sigma (\zeta )=-\left(\frac{2i\zeta }{\pi}\right) \int_0^\infty 
\frac{{\Re e}\ \sigma (\omega +i0^+)\ d\omega }{(\omega ^2-\zeta^2)}\ .
\label{tst5}
\end{eqnarray}
The frequency of plasma oscillations \begin{math} \omega_p  \end{math} 
may be found from the sum rule 
\begin{eqnarray}
4\pi \sigma (\zeta )\to \frac{i\omega_p^2}{\zeta }+\cdots 
\ \ \ \ (|\zeta|\to \infty),
\nonumber \\
\left(\frac{2}{\pi }\right)\int_0^\infty {\Re e}
\ \sigma (\omega +i0^+)\ d\omega 
=\frac{\omega_p^2}{4\pi }\ ,
\nonumber \\ 
\omega_p^2=4\pi e^2 \sum_a \left(\frac{z_a^2 n_a}{m_a}\right).
\label{tst6}
\end{eqnarray}
In Eq.(\ref{tst6}) the plasma component with charge 
\begin{math} z_a e \end{math} and mass \begin{math} m_a  \end{math} 
exists with a density of \begin{math} n_a \end{math} per unit volume. 
The thermodynamic velocity \begin{math} u_T \end{math} may be 
defined as 
\begin{equation}
u_T=\left(\frac{\omega_p}{\kappa }\right) .
\label{tst7} 
\end{equation}
The zero frequency conductivity \begin{math} \sigma \end{math}
defines a plasma relaxation time \begin{math} \tau \end{math} via 
\begin{math} 4\pi \sigma =\omega_p^2 \tau \end{math} or equivalently 
the charge diffusion coefficient \begin{math} D=u_T^2 \tau  \end{math} 
which in virtue of Eq.(\ref{tst7}) yields the Einstein relation 
\begin{equation}
4\pi \sigma =D \kappa ^2 .
\label{tst8}
\end{equation}
The above results are rigorously true for 
non-relativistic Coulomb plasma phases of matter. It is important to 
derive one further thermodynamic sum rule for the screening length 
\begin{math} \Lambda = \kappa^{-1}  \end{math}.

\subsection{Thermodynamic Sum Rules \label{tsr}}

The thermodynamic pressure 
\begin{math} P(T,\mu_1,\cdots,\mu_f) \end{math}
completely determines the equations of state of the 
plasma 
\begin{equation}
dP=sdT+\sum_a n_a d\mu_a .
\label{tsr1}
\end{equation}
Let us consider the charge in a macroscopic subvolume 
\begin{math} V \end{math} and the charge contained within 
that subvolume. Since the plasma is neutral, the mean charge 
is zero; i.e. 
\begin{equation}
\bar{Q}=e\sum_a z_a\bar{N}_a =0.
\label{tsr2}
\end{equation}
There are nevertheless charge fluctuations within the subvolume 
\begin{eqnarray}
\overline{\Delta Q^2}=e^2\sum_{a,b}
z_az_b\overline{\Delta N_a \Delta N_b }\ , 
\nonumber \\ 
\overline{\Delta Q^2}=k_BT
\left(\frac{\partial \bar{Q}}{\partial \Phi }\right)_T
=k_BT C_s,
\nonumber \\ 
\overline{\Delta N_a \Delta N_b }=Vk_BT
\left(\frac{\partial^2 P}{\partial \mu_a \partial \mu_b }\right)_T ,
\nonumber \\ 
\overline{\Delta Q^2}=Vk_BT e^2\sum_{a,b} z_a z_b
\left(\frac{\partial^2 P}{\partial \mu_a \partial \mu_b }\right)_T,  
\label{tsr3}
\end{eqnarray}
wherein \begin{math} \Phi \end{math} is a uniform electrostatic 
potential, the self capacitance of the volume \begin{math} V \end{math} 
is \begin{math} C_s \end{math} and statistical thermodynamic fluctuation 
theory\cite{fluctuation} has been invoked. Evidently, 
\begin{equation}
\frac{C_s}{V}=e^2\sum_{a,b} z_a z_b
\left(\frac{\partial^2 P}{\partial \mu_a \partial \mu_b }\right)_T .
\label{tsr4}
\end{equation}
The energy associated with a uniformly charge macroscopic subvolume 
\begin{math} V \end{math} obeys 
\begin{eqnarray}
{\cal E}=\frac{1}{2}\int \int 
\left[\frac{\rho_1\rho_2}{r_{12}}\right]{\cal S}(r_{12})
d^3{\bf r}_1 d^3{\bf r}_2 , 
\nonumber \\ 
{\cal E}=
\frac{\Delta Q^2}{2V}\int \frac{{\cal S}(r)}{r}d^3{\bf r}=
\frac{\Delta Q^2}{2C_s} \ ,  
\label{tsr5}
\end{eqnarray}
so that 
\begin{equation}
\frac{V}{C_s} = 4 \pi \int_0^\infty r{\cal S}(r)dr
=\frac{4\pi }{\kappa^2 }=4\pi \Lambda^2 
\label{tsr6}
\end{equation}
wherein Eq.(\ref{tst4}) has been invoked. In virtue of 
Eqs.(\ref{tsr4}) and (\ref{tsr6}) we have proved the following
\par \noindent 
{\bf Theorem:} The screening length 
\begin{math} \Lambda = \kappa ^{-1} \end{math}
is determined by the thermodynamic identity 
\begin{equation}
\kappa^2=4\pi e^2 \sum_{a,b} z_a z_b  
\left(\frac{\partial^2 P}{\partial \mu_a \partial \mu_b }\right)_T .
\label{tsr7}
\end{equation}
Depending on the equations of state implicit in Eq.(\ref{tsr1}), 
different screening lengths will appear in different regimes. For 
example, if the electrons are in a high temperature regime then 
the classical Debye screening length holds true. If the electrons 
are in a low temperature degenerate regime then the Thomas-Fermi 
screening length holds true. Let us consider the Debye screening 
regime.

\subsection{Debye Screening \label{ds}}

If the charged particles in the plasma are dilutely distributed 
then the particle number fluctuations 
\begin{eqnarray}
\overline{\Delta N_a \Delta N_b}=Vk_BT 
\left(\frac{\partial^2 P}{\partial \mu_a \partial \mu_b }\right)_T ,
\nonumber \\ 
\overline{\Delta N_a \Delta N_b}=\delta_{ab}\bar{N_a}
\ \ \ ({\rm dilute\ charged \ particles}).
\label{ds1}
\end{eqnarray}
Eqs.(\ref{tsr3}), (\ref{tsr7}) and (\ref{ds1}) imply the Debyre screening length 
\begin{math} \Lambda_D =\kappa_D^{-1} \end{math} is given by 
\begin{eqnarray}
\frac{1}{\Lambda_D^2}=\kappa_D^2
=\frac{4\pi e^2{\cal N}}{k_BT}
\ \ \ ({\rm Debye\ Screening}),
\nonumber \\ 
{\cal N}=\sum_a z_a^2 n_a = \frac{1}{L^3} 
\ \ \ ({\rm Ionicity}).
\label{ds2}
\end{eqnarray}
We will consider below electron proton scattering wherein the 
electron has a {\em heavy mass}  
\begin{math} \mu \end{math} with an effective Bohr radius 
\begin{equation}
a=\frac{\hbar^2}{\mu e^2}=
\frac{\hbar^2}{\beta m_e e^2}=\frac{a_B}{\beta } 
\label{ds3}
\end{equation}
wherein \begin{math} \beta \end{math} denotes the electron mass 
enhancement due to quickly oscillating electric and magnetic 
fields. The classical Debye screening Eq.(\ref{ds2}) is valid 
only in the regime\cite{Pitaevskii:2006} 
\begin{equation}
k_BT \gg \frac{e^2}{L} \gg \frac{e^2}{\Lambda_D}  
\gg \frac{e^2a}{L^2}\ ,  
\label{ds4}
\end{equation}
wherein the last inequality on the right hand side of 
Eq.(\ref{ds4}) requires that quantum corrections to Debye screening 
theory can be neglected. Debye screening itself is a purely 
classical effect. In the opposite regime that describes the quantum 
degeneracy, Thomas-Fermi screening is required. In all cases 
the central theorem Eq.(\ref{tsr7}) for screening lengths holds true. 
Finally, the Debye theory for the thermal velocity 
\begin{math} u_T \end{math} in Eq.(\ref{tst7}) yields  
\begin{eqnarray}
\frac{1}{m}=\frac{\sum_a z_a^2 n_a /m_a}{\sum_a z_a^2 n_a}\ ,
\nonumber \\ 
u_T=\sqrt{\frac{k_BT}{m}}\ ,
\label{ds5}
\end{eqnarray}
that again indicates the classical nature of the Debye screening theory.

\section{Electron Correlations \label{ec}}

Consider the following electron capture process in a nucleus in the 
vacuum,
\begin{equation}
e^- +\  ^A_Z X \to \nu_e +\  ^A_{Z-1} X.  
\label{ec1}
\end{equation}
One can employ a complex scattering length 
\begin{math} f=\tilde{a}+ib \end{math} in the center of inertia 
reference frame to describe Eq.(\ref{ec1}) in the low relative 
velocity limit; The elastic and total cross sections are thereby 
\begin{eqnarray}
\sigma_{el}=4\pi |f|^2=4\pi (\tilde{a}^2+b^2),
\nonumber  \\ 
\sigma_{tot}(v)=\left(\frac{4\pi }{k}\right){\Im m}f
=\frac{4\pi \hbar b}{\mu v}\ \ \ \ \ (v\to 0). 
\label{ec2}
\end{eqnarray}
When the nuclei is embedded in a condensed matter plasma, the 
transition rate per nucleus obeys  
\begin{equation}
\Gamma=\lim_{v\to 0}\left[v\sigma_{tot}(v)\right]\bar{n}=
\left(\frac{4\pi \hbar b}{\mu }\right)\bar{n},   
\label{ec3}
\end{equation}
wherein \begin{math} \bar{n}  \end{math} is the density of electrons 
at positions \begin{math} \{{\bf r}_j\} \end{math} that reside right 
on top of the nucleus at position 
\begin{math} {\bf R} \end{math}; It is the electron nuclear 
correlation function
\begin{equation}
\bar{n}=\left<\sum_j \delta ({\bf R}-{\bf r}_j) \right>   
\label{ec4}
\end{equation}
For the case of a heavy electron \begin{math} \tilde{e}^- \end{math}
dressed in a cloud of photons scattering off a proton producing a 
neutrino and a neutron  
\begin{equation}
\tilde{e}^- + p^+ \to \nu_e +n   
\label{ec5}
\end{equation}
we wish to evaluate the correlation function in Eq.(\ref{ec4}) in order 
to describe the effects of the plasma on the rate for Eq.(\ref{ec5}).

\subsection{Numerical Length Scales \label{ls}}

\begin{table}
\caption{Numerical Values of Length Scales}
\label{tab1}
\begin{center}
\begin{tabular}{lr}
	\hline 
		$\beta $ & \ \ \ \ \ $20 $ \\ 
		$T$ & \ \ \ \ \ $5.0 \times 10^3\ ^o$K \\ 
		$a$ & \ \ \ \ \ $2.65 \times 10^{-10} $\ cm \\ 
		$l_T$ & \ \ \ \ \ $3.34 \times 10^{-7} $\ cm \\ 
		$L$ & \ \ \ \ \ $5.0 \times 10^{-5} $\ cm \\
		$\Lambda_D$ & \ \ \ \ \ $1.73 \times 10^{-4} $\ cm \\
		$\lambda_T$ & \ \ \ \ \ $ 2.36\times 10^{-8} $\ cm \\
	\hline 
\end{tabular}
\par\medskip\footnotesize
Order of magnitude estimates are given for a chemical cell 
water plasma. Eqs.(\ref{ds4}) and (\ref{ls7}) for Debye theory 
validity holds true.
\end{center}
\end{table}

There are many length scales that need to be considered in the 
evaluation of the electron correlation function Eq.(\ref{ec4}) 
for the reaction Eq.(\ref{ec5}). (i) The effective Bohr radius 
is given by  
\begin{equation}
a=\frac{\hbar^2}{\mu e^2}=\frac{a_B}{\beta } 
\ \ \ \ \ \ \ \ \ \ \ \ ({\rm Effective\ Bohr\ Radius}).
\label{ls1}
\end{equation}
(ii) The Landau length is 
\begin{equation}
l_T=\frac{e^2}{k_BT}
\ \ \ \ \ \ \ \ \ \ \ \ \ \ \ \ \ \ \ \  
\ \ \ \ \ \ ({\rm Landau\ Length}).
\label{ls2}
\end{equation}
Let \begin{math} n \end{math} represents the number density of 
electrons which is the same as the number of protons in the 
neutral plasma. (iii) The mean particle spacing is given by 
\begin{equation}
L=n^{-1/3}
\ \ \ \ \ \ \ \ \ \ \ \     
\ \ \ \ \ ({\rm Mean\ Particle\ Spacing}).
\label{ls3}
\end{equation}
(iv) The Debye screening length is  
\begin{equation}
\Lambda_D=L\sqrt{\frac{L}{4\pi l_T}}
\ \ \ \ \ \ \ \ ({\rm Debye\ Screening\ Length}).
\label{ls4}
\end{equation}
(v) The thermal De Broglie wave length is 
\begin{equation}
\lambda_T=\sqrt{\frac{2\pi \hbar^2}{\mu k_BT}}
\ \ \ \ \ \ \ \ ({\rm De\ Broglie\ Wave\ Length}).
\label{ls5}
\end{equation}
Equivalently, the thermal De Broglie wave length is 
determined by 
\begin{eqnarray}
\int e^{-p^2/2\mu k_BT} 
\left[\frac{d^3{\bf p}}{(2\pi \hbar)^3}\right]
=\frac{1}{\lambda_T^3}\ ,
\nonumber \\ 
\lambda_T=\sqrt{\frac{2\pi \hbar^2}{\mu k_BT}}=
\sqrt{2\pi a l_T} \ . 
\label{ls6}
\end{eqnarray}
A list of order of magnitude estimates for the above length scales 
in a chemical cell exhibiting nuclear transmutations is given in 
TABLE \ref{tab1}. Eq.(\ref{ds4}) for the validity of Debye theory may be 
expressed as 
\begin{equation}
l_T \ll L \ll \Lambda_D \ll L^2/a\ , 
\label{ls7}
\end{equation}in agreement with the estimates in TABLE \ref{tab1}. On 
the other Maiani {\em et. al.}\cite{Maiani:2014} has erroneously 
applied the Debye screening formula in regimes wherein 
Eq.(\ref{ls7}) is violated by a large margin.  

\subsection{Application to Electron Capture \label{aec}}

In the limit of a small thermal De Broglie wave length  
\begin{eqnarray}
\lambda_T\ll \Lambda_D\ ,  
\nonumber \\ 
\bar{n}\approx n\left<\frac{2\pi e^2}{\hbar v}\right>_T  
\approx \sqrt{\frac{8\pi \mu }{k_BT}}\left(\frac{e^2 n}{\hbar }\right),
\nonumber \\ 
\bar{n}\approx 2\left(\frac{l_T}{\lambda_T}\right)n 
\approx \left(\frac{1}{2\pi \lambda_T \Lambda_D^2}\right) . 
\label{aec2}
\end{eqnarray}
Eqs.(\ref{aec2}) is a hot Debye screening plasma result equivalent to 
our previous calculations\cite{widom:2112,widom:2113}.

\section{s-State Wavefunctions \label{ssw}}

Employing the definitions 
\begin{equation}
E=\frac{\hbar^2 k^2}{2\mu }\ \ \ {\rm and}
\ \ \ \phi (r)=\frac{2\mu {\cal U}(r)}{\hbar^2 }\ ,
\label{ssw1}
\end{equation}
the s-wave radial wave function 
\begin{equation}
\psi (r)=\frac{u(r)}{r}\ ,
\label{ssw2}
\end{equation}
is governed by the potential scattering Eq.(\ref{intro2}), 
\begin{equation}
u^{\prime \prime }(r,k)+\big(k^2-\phi(r)\big)u(r,k)=0.
\label{ssw3}
\end{equation}
Calogero\cite{Calogero:1967} defines a variable phase 
\begin{math} \eta(r,k) \end{math} and variable amplitude 
\begin{math} w(r,k) \end{math} defined by  
\begin{eqnarray}
u(r,k)=w(r,k)\sin\big(kr+\eta(r,k)\big), 
\nonumber \\ 
u^\prime (r,k)=kw(r,k)\cos\big(kr+\eta(r,k)\big). 
\label{ssw4}
\end{eqnarray}
The s-wave phase shift \begin{math} \eta_s(k) \end{math} is 
computed in virtue of the limits 
\begin{eqnarray}
\lim_{r\to 0}\eta (r,k)=0, 
\nonumber \\ 
\lim_{r\to \infty }\eta (r,k)=\eta_s(k). 
\label{ssw5}
\end{eqnarray}
The second order differential Eq.(\ref{ssw3}) is thereby 
replaced by two first order differential equations 
\begin{equation}
\eta^\prime (r,k)=-\left(\frac{\phi (r)}{k}\right)
\sin^2 \big(kr+\eta (r,k)\big), 
\label{ssw6}
\end{equation} 
and 
\begin{equation} 
w^\prime (r,k)=\left(\frac{\phi(r)}{2k}\right)
w(r,k) \sin\big(2kr+2\eta (r,k)\big). 
\label{ssw7}
\end{equation}
Eqs.(\ref{ssw5}) and (\ref{ssw6}) determine the variable 
phase shift which which in turn determines the variable 
amplitude in virtue of Eq.(\ref{ssw7}); i.e. 
\begin{eqnarray} 
w(r,k)=w(0,k) \times 
\nonumber \\ 
\exp\left[\int_0^r 
\left(\frac{\phi(r^\prime )}{2k}\right)
\sin\big(2kr^\prime +
2\eta (r^\prime ,k)\big) dr^\prime \right]. 
\label{ssw8}
\end{eqnarray}
Of interest in what follows is the limit 
\begin{math} k\to 0 \end{math}. This small energy regime 
is described by the variable scattering length. 

\subsection{Variable Scattering Length \label{vsl}}

The scattering length \begin{math} {\cal L}_s \end{math}
is here defined in terms of the s-wave phase shift as 
\begin{equation}
{\cal F}_s = \lim_{k\to 0} \left(\frac{\eta_s(k)}{k}\right).
\label{vsl1}
\end{equation} 
The variable scattering length is defined in terms of the variable 
phase according to   
\begin{equation}
{\cal F}(r) = \lim_{k\to 0}\left( \frac{\eta(r,k)}{k} \right).
\label{vsl2}
\end{equation} 
so that 
\begin{equation}
{\cal F}_s = \lim_{r\to \infty }{\cal F}(r). 
\label{vsl3}
\end{equation}
In virtue of Eqs.(\ref{ssw6}) and (\ref{vsl2}), one finds 
\begin{eqnarray}
{\cal F}(0)=0,
\nonumber \\
{\cal F}^\prime (r)= -\phi(r) \big(r+{\cal F}(r) \big)^2 , 
\nonumber \\ 
{\cal F}_s=-\int_0^\infty \phi(r) \big(r+{\cal F}(r) \big)^2 dr.   
\label{vsl4}
\end{eqnarray}
The s-wave cross section is thereby 
\begin{equation}
\lim_{k\to 0} \sigma_s(k)=4\pi |{\cal F}_s|^2.
\label{vsl5}
\end{equation}
Finally, the relative probability, i.e. relative amplitude 
squared, for the scattering particles to be on top of one 
another compared with being widely separated 
\begin{eqnarray}
\lim_{k\to 0}\left|\frac{w(0,k)}{w(\infty ,k)} \right|^2=
\nonumber \\ 
\exp \left(
-2\int_0^\infty \phi(r) \big(r+{\cal F}(r)\big) dr 
\right).
\label{vsl6}
\end{eqnarray}
Eqs.(\ref{vsl4})  and (\ref{vsl6}) is central for predicting 
electron capture rates in the degenerate Thomas-Fermi screening regime.

\subsection{Degenerate Electron Capture \label{dec}}

Suppose we consider an attractive screened Coulomb potential of the form 
\begin{equation}
{\cal U}(r)=-\left[\frac{|Z_1Z_2| e^2}{r}\right] e^{-(r/\Lambda )} ,  
\label{dec1}
\end{equation}
wherein the screening length \begin{math} \Lambda  \end{math} is not 
required to be a Debye screening length. Then 
\begin{math} \phi (r) \end{math} has the form 
\begin{equation}
\phi(r)=-\left[\frac{2}{ar}\right] e^{-(r/\Lambda )}.  
\label{dec2}
\end{equation}
Eqs.(\ref{vsl4}) and (\ref{dec2}) imply 
\begin{equation}
{\cal F}^\prime (r)=\left[\frac{2}{ar}\right] e^{-(r/\Lambda )}
\big(r+{\cal F}(r)\big)^2 \ \ \ \Rightarrow 
\ \ \ {\cal F}(r)\ge 0.  
\label{dec3}
\end{equation}
The scattering length thereby obeys 
\begin{eqnarray}
{\cal F}_s = \frac{2}{a}
\int_0^\infty \frac{1}{r}\big(r+{\cal F}(r) \big)^2 
e^{-(r/\Lambda )}dr, 
\nonumber \\ 
{\cal F}_s \ge 
\frac{2}{a}\int_0^\infty r e^{-(r/\Lambda )}dr 
= \frac{2\Lambda^2 }{a}.
\label{dec4}
\end{eqnarray}
The cross section is bounded from below by 
\begin{equation}
\lim_{k\to 0} \sigma_s(k) = 4\pi |{\cal F}_s|^2
\ge \left(\frac{16\pi \Lambda^4}{a^2}\right).
\label{dec5}
\end{equation}
As the screening length grows ever larger 
\begin{math} \Lambda \to \infty  \end{math} 
the scattering cross section also diverges; i.e. 
the total cross section of an unscreened Coulomb potential 
is infinite.

The rate of electron capture for this model is determined by 
Eqs.(\ref{ec3}) and (\ref{ec4}) via 
\begin{eqnarray}
\bar{n}=n\lim_{k\to 0}\left|\frac{w(0,k)}{w(\infty ,k)} \right|^2, 
\nonumber \\ 
\bar{n}=n \exp \left(
-2\int_0^\infty \phi(r) \big(r+{\cal F}(r)\big) dr 
\right),
\label{dec6}
\end{eqnarray}
in virtue of Eq.(\ref{vsl6}). 
From Eqs.(\ref{dec2}), (\ref{dec3}) and (\ref{dec6}), 
\begin{eqnarray}
\bar{n}\ge n \exp
\left( \frac{4}{a}\int_0^\infty e^{-(r/\Lambda )}dr \right), 
\nonumber \\ 
\bar{n}\ge n \exp \left(\frac{4\Lambda }{a}\right).
\label{dec7}
\end{eqnarray}
The lower bound in Eq.(\ref{dec7}) is exponentially larger than the 
prediction of Maiani {\em et. al.}\cite{Maiani:2014} for the quantum 
degenerate density \begin{math} n \end{math} discussed in this 
section.

\section{Conclusion \label{conc}}

Properties of a fully ionized water plasma have been discussed.  
The theory of the screening of the Coulomb law has been rigorously 
derived from a thermodynamic viewpoint. A kinetic model was reviewed 
determining the transition rate per unit time for electron capture by 
a nucleus and the resulting nuclear transmutations. 
Corrections to the recent Maiani {\em et.al.} calculations have been 
discussed. The regime of validity for Debye screening length has been 
derived. The WKB approximation erroneously employed by Maiani in the long 
De Broglie wave was corrected employing the mathematically rigorous Calogero 
formalism in potential scattering. We have stood by our previous results 
on the rates of electro electron capture processes for a water plasma in 
chemical cells.

\end{document}